\documentclass{aa}  
\usepackage{graphicx}
\usepackage{txfonts}
\begin{document}

\title{Observations and analysis of two type IIP supernovae: the intrinsically 
 faint object SN 2005cs and the ambiguous object SN 2005ay}

\titlerunning{SNe 2005ay \& 2005cs}  

  \author{D.Yu. Tsvetkov\inst{1}, A.A. Volnova\inst{1}, A.P. Shulga\inst{1},
         S.A. Korotkiy\inst{2}, A. Elmhamdi \inst{3,4}, I.J. Danziger \inst{4},
          M.V. Ereshko\inst{2}}

   \offprints{D. Tsvetkov \& A. Elmhamdi, {tsvetkov@sai.msu.su; elmhamdi@ts.astro.it}}
   
   \institute{Sternberg State Astronomical Institute, Universitetskii pr. 13, 
              119992, Moscow, Russia
         \and SC "Ka-Dar", Neopalimovskii 1st per. 16/13, 119121, 
           Moscow, Russia
         \and ICTP-International Centre for Theoretical Physics, Strada Costiera 11, 34014 
         Trieste - Italy
	 \and INAF-Osservatorio Astronomico di Trieste, Via G.B.Tiepolo 11 - I-34131
	 Trieste - Italy 
}
 
\authorrunning{Tsvetkov et al.}

\date{}
\abstract{}
{ To derive observational properties and physical parameters of the 
 progenitor stars of type IIP supernovae SN 2005ay and SN 2005cs from
 their $U,B,V,R,I$ CCD photometry, and to define 
 their velocity behaviour.}
 {Light curves are analysed, and the velocities
 and spectral characteristics of SN 2005cs are  
 obtained using synthetic spectra modeling.}
 {Both supernovae are found to be fainter than the average
 SN IIP, with SN 2005cs being more subluminous and showing slight  
 brightening in the second half of plateau stage in the
 $V,R,I$ bands and a low expansion velocity.
 The effects of two different plausible distance moduli on the
 derived physical parameters of SN 2005ay are considered.
 Two approaches are used to recover the amounts of the 
 ejected $^{56}$Ni, indicating masses of the order of 
 $\sim$ 0.02 M$_\odot$, although late luminosities might indicate
 a higher amount for SN 2005ay, especially for the large distance 
 case.
 Constraints on the progenitor properties are also presented, based
 on empirical analytical models. Two approaches are used
  to estimate the expansion velocities at the middle of
 the plateau phase. SN 2005cs represents an example of where all 3 physical
 parameters, velocity, energy and $^{56}$Ni mass are lower than
 average, a correlation not always observed in SNe IIP. SN 2005ay
 may belong to the same class if the shorter distance possibility
 is adopted. 
 Furthermore, the estimated mass range for SN 2005cs is in agreement
 with limits established by using pre-supernova imaging.}
{} 
\keywords{Supernovae: type IIP, spectra, light curves; Individual: SN 2005ay, SN2005cs}
\maketitle

\section{Introduction}

In 2005 two bright type II supernovae (SN) were discovered
in nearby spiral galaxies: SN 2005ay in NGC 3938 and SN 2005cs
in M 51 (NGC 5194). 
\medskip

SN 2005ay was discovered on March 27 by D. Rich at 
magnitude 15.6 on CCD frames taken with a 0.31-m 
reflector (IAUC 8500)\nocite{Rich05}. 
The SN was located at $\alpha = 11^{\rm h}52^{\rm m}48^{\rm s}.07,  
\delta = +44^\circ06'18''.4$ (equinox 2000.0), which is 
$15''$ west and $56''$ south of the center of Sc galaxy
NGC 3938. 
Taubenberger et al. (2005)\nocite{Tau05} reported that a spectrum
of SN 2005ay, obtained on March 29.98 UT with the Calar Alto 2.2-m telescope 
was that of a type-II SN soon (perhaps 8-10 days) after
explosion. The spectrum was characterized by a blue continuum with
broad P-Cyg features of the hydrogen Balmer series (the expansion
velocity from H$\beta$ was about 9600 km s$^{-1}$). 

\medskip

The discovery of SN 2005cs was reported by W. Kloehr
on June 28 at a magnitude of about 14 (IAUC 8553)\nocite{Klo05}.
W. Li confirmed the discovery and provided the position of the SN
as  $\alpha = 13^{\rm h}29^{\rm m}53^{\rm s}.37,
\delta = +47^\circ10'28''.2$ (equinox 2000.0), which is
$15''$ west and $67''$ south of the center of Sbc galaxy M51 
 (IAUC 8553)\nocite{Li05}.
Nothing was visible at this location on earlier frames 
 taken by W. Kloehr on May 11 and 26. A KAIT image taken on June 15.20 showed 
 nothing at this position (limiting mag about 19.0), nor on 
 Itagaki's image taken in poor conditions on
June 20.598 (limiting mag 17.0).

Modjaz et al. (2005)\nocite{Mod05} reported that a spectrogram 
of SN 2005cs obtained on June
30.23 UT with the F. L. Whipple Observatory 1.5-m telescope 
showed it to be a young type-II SN.  The spectrum
consisted of a blue continuum and P-Cyg profiles of the Balmer and
He lines. The expansion
velocity derived from the minimum of the H$\beta$ line was about 7500
km s$^{-1}$. Zero-velocity interstellar Na I D absorption with an
equivalent width of about 0.02 nm was detected in the spectrum of SN
2005cs, indicating some reddening by dust. 
Interstellar Na I D absorption was  
detected also at the redshift of the host galaxy, with comparable
equivalent width, suggesting some host-galaxy extinction.
\begin{table*}
\centering
\caption{Colour terms for reduction equations}\vskip2mm
\begin{tabular}{lcccccc}
\hline
Telescope, & ¤$K_u$ & $K_b$ & $K_v$ & $K_r$ & $K_i$ \\
CCD, filters &      &       &       &       &       \\
\hline
C60a1 & -0.03 & -0.21 & -0.017 &  0.09  & -0.38 \\
C60a2 & -0.03 & -0.21 & -0.017 &  -0.45  & -0.38 \\
C60c  & -0.06 & -0.10 & -0.026 &  -0.025 & -0.43 \\
M70a1 & -0.05 & -0.21 & -0.023 &   0.09  & -0.40 \\
M70a2 & -0.05 & -0.09 & -0.035 & -0.10 & -0.38 \\
M70b1 &       & -0.14 & -0.023 &  -0.12  & -0.38 \\
M70b2 &       & -0.14 & -0.023 &  -0.45  & -0.38 \\         
\hline
\end{tabular}
\end{table*}

\begin{table*}
\caption{Magnitudes of comparison stars}\vskip2mm
\begin{tabular}{lcccccccccc}
\hline
Star & $U$ & $\sigma_U$ & $B$ & $\sigma_B$ & $V$ & $\sigma_V$ & $R$ & $\sigma_R$ & $I$ & $\sigma_I$  \\
\hline 
NGC3938-1 &14.47 & 0.05 & 14.40 & 0.03 & 13.83 & 0.02 & 13.49 & 0.02 & 13.17 & 0.02 \\
NGC3938-2 & 16.36 & 0.08 & 15.86 & 0.03 & 15.07 & 0.02 & 14.63 & 0.02 & 14.25 & 0.02 \\
M51-1 &14.19 &0.04 &14.03 & 0.01 & 13.41 & 0.01 & 13.07 & 0.01 & 12.76 & 0.02 \\
M51-2 &      &   & 16.42 & 0.05 & 15.11 & 0.02 & 14.35 & 0.01 & 13.71 & 0.02 \\ 
\hline
\end{tabular}
\end{table*}

\begin{figure}
\includegraphics[height=9cm,width=9cm]{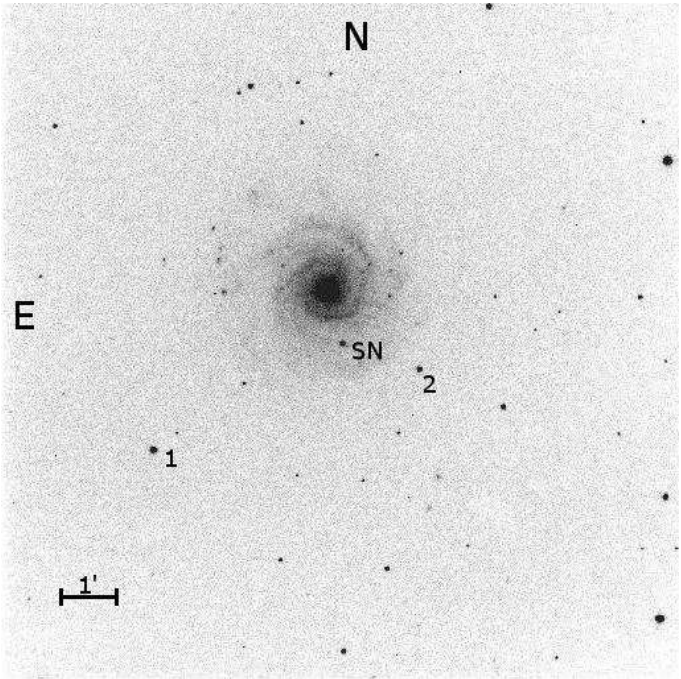}
\caption{SN 2005ay in NGC~3938 with comparison stars }
\end{figure}
\begin{figure}
\includegraphics[height=9cm,width=9cm]{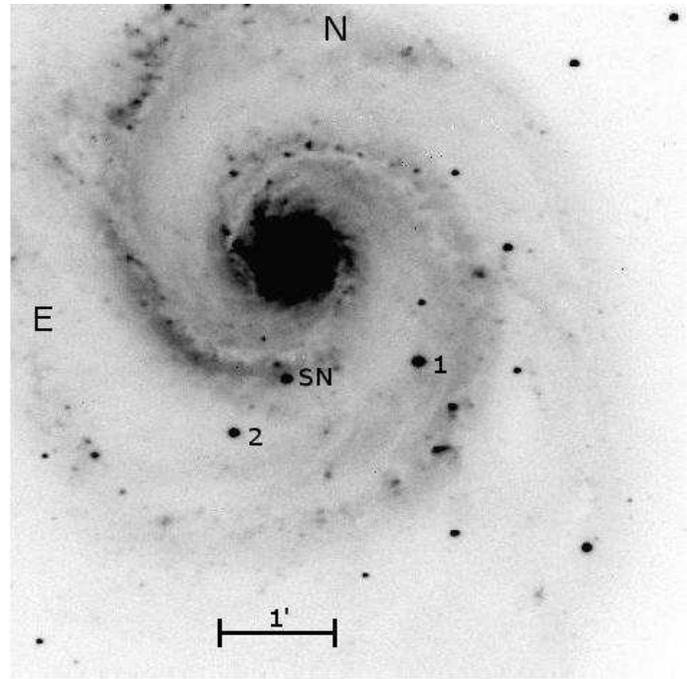}
\caption{SN 2005cs in M~51 with comparison stars}   
\end{figure}

Maund et al. (2005)\nocite{Mau05} and Li et al. (2006)\nocite{Li06} 
reported detection of the progenitor of SN 2005cs
in pre-explosion archival images obtained with the Hubble Space
Telescope. They concluded that it was a red supergiant of
spectral type K0-M3, with initial mass $M_{ZAMS}=9^{+3}_{-2}~ M_\odot$. 
 Li et al. also provided a provisional light curve for SN 2005cs
and supposed that it was a subluminous type IIP event.  
\section{Observations and reductions}
The photometric observations were carried on with the 
70-cm reflector of the Sternberg Astronomical Institute in Moscow (M70)
using Apogee AP-47p (a) and AP-7p (b) CCD cameras, and with the
  60-cm reflector of the Crimean 
Observatory of the Sternberg Astronomical Institute (C60) equipped with 
a Roper Scientific VersArray1300B (c) and Apogee AP-47p cameras.
On July 14 observations were obtained at the 1-m telescope of
 the Special Astrophysical Observatory (S100) with CCD EEV42-40,
and on July 11 at the 50-cm meniscus telescope in the Crimea (C50) with 
 a CCD Pictor416.

All reductions and photometry were made using IRAF\footnote{IRAF is 
distributed by the National Optical Astronomy Observatory,
which is operated by AURA under cooperative agreement with the
National Science Foundation}.

The colour terms for transformation of instrumental magnitudes $ubvri$
to standard $UBVR_cI_c$ were determined for different
telescope-filter-detector
combinations using observations of standards in M~67 (Chevalier $\&$ 
 Ilovaisky 1991)\nocite{Chev91} and in NGC~7790 (Stetson 2000)\nocite{Stet00}. 
The equations $u=U+K_u(U-B)+C_u; b=B+K_b(B-V)+C_b; v=V+K_v(B-V)+C_v; 
r=R+K_r(V-R)+C_r; i=I+K_i(R-I)+C_i$ were solved for colour terms,
which are listed in Table 1.
The codes for telescopes and CCD cameras are followed by numbers denoting
the filter sets.
The colour terms for S100 were kindly given to us by V.P.Goranskij.
Observations with the C50 were obtained only in the $V$ filter which
was close to the standard system, and no correction was applied.
\begin{table*}
\caption{$BVRI$ photometry of SN 2005ay}\vskip2mm
\begin{tabular}{cccccccccl}
\hline
JD 2453000+ & $B$ & $\sigma_B$ & $V$ & $\sigma_V$ & $R$ & $\sigma_R$ & 
$I$ & $\sigma_I$ & Tel.\\
\hline
458.45& 15.32& 0.09 &  15.12& 0.08&  14.98& 0.05&  15.02& 0.09 & M70a1 \\
463.36& 15.46& 0.08 &  15.19& 0.05&  14.95& 0.05&  14.95& 0.04 & M70a1 \\
465.42& 15.47& 0.06 &  15.20& 0.06&  14.94& 0.07&  14.94& 0.05 & M70a1 \\
468.40& 15.60& 0.10 &  15.21& 0.05&  14.97& 0.08&  14.81& 0.07 & M70a1 \\
473.40& 15.76& 0.06 &  15.16& 0.04&  14.94& 0.04&  14.80& 0.05 & M70a1 \\
494.38& 16.50& 0.09 &  15.54& 0.06&  15.06& 0.06&  14.85& 0.16 & M70b2 \\
495.32& 16.37& 0.07 &  15.33& 0.06&  15.04& 0.04&  14.70& 0.12 & C60c  \\
500.36& 16.60& 0.06 &  15.40& 0.04&  15.15& 0.04&  14.75& 0.06 & C60c  \\
506.41& 16.53& 0.08 &  15.45& 0.06&  15.02& 0.06&  14.74& 0.11 & M70b2 \\
509.38& 16.67& 0.09 &  15.43& 0.10&  14.99& 0.06&  14.73& 0.09 & M70b1 \\
515.37& 16.67& 0.08 &  15.42& 0.10&  15.03& 0.07&  14.73& 0.11 & M70b1 \\
560.34& 17.65& 0.11 &  16.33& 0.04&  15.75& 0.04&       &      & C60a2 \\
562.31& 18.10& 0.14 &  16.56& 0.04&  15.83& 0.03&  15.44& 0.10 & C60a2 \\
566.30&      &      &  16.98& 0.08&  16.30& 0.03&  15.80& 0.11 & C60a2 \\
566.31& 18.66& 0.05 &  17.14& 0.04&  16.42& 0.03&  15.88& 0.04 & S100 \\
569.29& 18.61& 0.14 &  17.37& 0.10&  16.58& 0.05&  16.33& 0.15 & C60a2 \\
680.63&      &      &  18.67& 0.08&  17.71& 0.05&       &      & C60a1 \\
683.58& 19.90& 0.10 &  18.76& 0.06&  17.81& 0.04&  17.62& 0.11 & C60a1 \\
687.57& 19.83& 0.15 &  18.70& 0.06&  17.81& 0.09&       &      & C60a1 \\
688.55&      &      &  18.85& 0.16&  17.74& 0.12&       &      & C60a1 \\
804.56&      &      &       &     &  18.83& 0.28&       &      & M70a1 \\
852.37&      &      &       &     &  19.36& 0.24&       &      & M70a2 \\
\hline
\end{tabular}
\end{table*}

\begin{table*}
\caption{Observations of SN 2005cs}\vskip2mm
\begin{tabular}{cccccccccl}
\hline
JD 2453000+ & $B$ & $\sigma_B$ & $V$ & $\sigma_V$ & $R$ & $\sigma_R$ &
$I$ & $\sigma_I$ & Tel.\\
\hline
555.33 & 14.45 & 0.02 & 14.45 & 0.02 & 14.30 & 0.02 & 14.37 & 0.05 & C60a2\\
558.32 & 14.57 & 0.02 & 14.54 & 0.02 & 14.34 & 0.02 & 14.29 & 0.03 & C60a2\\
561.32 & 14.74 & 0.02 & 14.50 & 0.02 & 14.26 & 0.02 & 14.25 & 0.03 & C60a2\\
563.30 &       &      & 14.50 & 0.02 &       &      &       &      & C50 \\
564.29 & 14.92 & 0.02 & 14.50 & 0.02 & 14.32 & 0.02 & 14.20 & 0.03 & C60a2\\
565.36 & 15.08 & 0.05 & 14.65 & 0.04 & 14.42 & 0.04 & 14.29 & 0.06 & M70b1\\
566.33 & 15.19 & 0.02 & 14.63 & 0.02 & 14.39 & 0.02 & 14.32 & 0.03 & S100 \\
567.29 & 15.13 & 0.03 & 14.60 & 0.02 & 14.34 & 0.02 & 14.23 & 0.03 & C60a2\\
571.29 & 15.40 & 0.02 & 14.72 & 0.02 & 14.43 & 0.02 & 14.27 & 0.03 & C60a2\\
571.40 & 15.34 & 0.02 & 14.72 & 0.02 & 14.38 & 0.04 & 14.22 & 0.04 & M70b1\\
580.32 & 15.69 & 0.05 & 14.73 & 0.03 & 14.42 & 0.03 & 14.20 & 0.04 & M70b1\\
587.31 & 15.77 & 0.05 & 14.72 & 0.03 & 14.31 & 0.02 & 14.08 & 0.03 & M70b1\\
593.31 & 15.82 & 0.03 & 14.72 & 0.03 & 14.28 & 0.02 & 14.03 & 0.03 & M70b1\\
607.34 & 15.89 & 0.06 & 14.63 & 0.12 & 14.15 & 0.07 & 13.89 & 0.08 & M70b1\\
612.29 & 16.08 & 0.03 & 14.74 & 0.02 & 14.14 & 0.02 & 13.87 & 0.05 & C60a1\\
613.36 &       &      & 14.63 & 0.06 & 14.14 & 0.05 & 13.84 & 0.13 & C60a1\\
618.28 & 15.99 & 0.06 & 14.71 & 0.05 & 14.16 & 0.03 &       &      & C60c\\
620.25 &       &      & 14.68 & 0.03 & 14.12 & 0.02 & 13.83 & 0.04 & C60c\\
621.25 & 16.15 & 0.08 & 14.84 & 0.07 & 14.15 & 0.03 & 13.80 & 0.06 & M70b1\\
628.25 & 16.11 & 0.04 & 14.79 & 0.04 & 14.17 & 0.04 & 13.83 & 0.05 & M70b1\\
644.19 & 16.53 & 0.19 & 14.96 & 0.11 & 14.29 & 0.05 & 13.96 & 0.10 & M70b1\\
653.24 & 16.37 & 0.13 & 15.04 & 0.15 & 14.39 & 0.08 & 14.03 & 0.12 & M70b1\\
683.63 &       &      & 17.91 & 0.15 & 17.20 & 0.11 & 16.58 & 0.09 & C60a1\\
687.60 &       &      & 17.77 & 0.19 & 17.22 & 0.14 &       &      & C60a1\\
804.56 &       &      & 18.62 & 0.18 & 17.81 & 0.11 &       &      & M70a2\\
831.43 &       &      & 18.41 & 0.11 & 18.07 & 0.10 &       &      & M70a2\\
850.46 &       &      & 18.59 & 0.15 & 18.30 & 0.09 & 17.62 & 0.14 & M70a1\\
\hline
\end{tabular}
\end{table*}

\begin{table*}
\caption{Observations of SNe in the $U$ band}\vskip2mm
\begin{tabular}{ccclcccl}
\hline
\multicolumn{4}{c}{SN 2005ay} & \multicolumn{4}{c}{SN 2005cs}\\
\hline
JD 2453000+ & $U$ & $\sigma_U$ & Tel.& JD 2453000+ &$U$&$\sigma_U$ & Tel.\\
\hline
458.450 &14.67& 0.11 & M70a1  & 566.33 & 15.36  & 0.06  & S100  \\
463.360 &14.84& 0.21 & M70a1  &  &  &  &  \\
465.420 &15.01& 0.20 & M70a1  &  &  &  &  \\
495.320 &17.58& 0.16 & C60c   &  &  &  &  \\
500.360 &17.57& 0.10 & C60c   &  &  &  &  \\
\hline
\end{tabular}
\end{table*}

The images of the SNe with comparison stars are shown in Figs. 1,2.
The magnitudes of comparison stars were determined on 
photometric nights, when we observed standards from Landolt 
(1992)\nocite{Land92} and standard regions in the clusters M~67 
and M~92\footnote{ 
http://cadcwww.hia.nrc.ca/cadcbin/wdb/astrocat/stetson/query/
NGC6341}.
The magnitudes of stars M~51-1 and M~51-2 were derived by
Richmond et al. (1996)\nocite{Rich96}. We included their data in 
 the calculation of mean values, which are presented in Table 2 
 with their RMS errors. 

Photometric measurements of the SNe were made relative to 
comparison stars using PSF-fitting with the IRAF DAOPHOT package,
and in some cases using aperture photometry with the APPHOT package.
The background of the host galaxies around the SNe 
did not present any problems when the SNe were bright, but at late stages
it could introduce additional errors. As the SNe are still visible on our
images, we cannot apply image subtraction. Thus, the magnitudes
at late stages can be regarded as provisional, and they should be
verified later when images without the SNe are obtained and
used for image subtraction. 
The results of SNe observations are presented in Tables 3,4,5.

Spectra of SN 2005cs were obtained on July 6 at the S100 with 
 the UAGS spectrograph in the range
3425-6200\AA\  with a dispersion of 1.34\AA\  per pixel.
The spectra were extracted and wavelength and
flux calibrated using the IRAF KPNOSLIT algorithm. The wavelength 
 calibration 
was done using exposures of Ne lamp made before and after target
observations. For flux calibration BD+28$^\circ$4211 was observed
as a flux standard.
\begin{table*}
\caption{Main parameters of the SNe}\vskip2mm
\begin{tabular}{ccccccc}
\hline
SN & Host & Explosion & Distance & $E(B-V)$ & M$_V$$^1$ & \\
name  & Galaxy & time (JD) & modulus & (total) & at $\sim$300d &\\
\hline
2005ay &NGC 3938 & 2453453 & 30.82$-$31.71  & 0.1 & -11.78 $-$ -12.67  \\
2005cs &M 51 & 245348.8 & 29.62  & 0.14 & -11.46   \\
\hline
\end{tabular}\\
\footnotesize
\hskip-5.5cm\emph{\rm 1- Extrapolated at day 300 for comparison. \\
$~~~$ Estimates adopting short and large distances are reported for SN 2005ay}\\ 
\normalsize
\end{table*}
\section{The analysis}
\begin{figure}
\includegraphics[height=9cm,width=9cm]{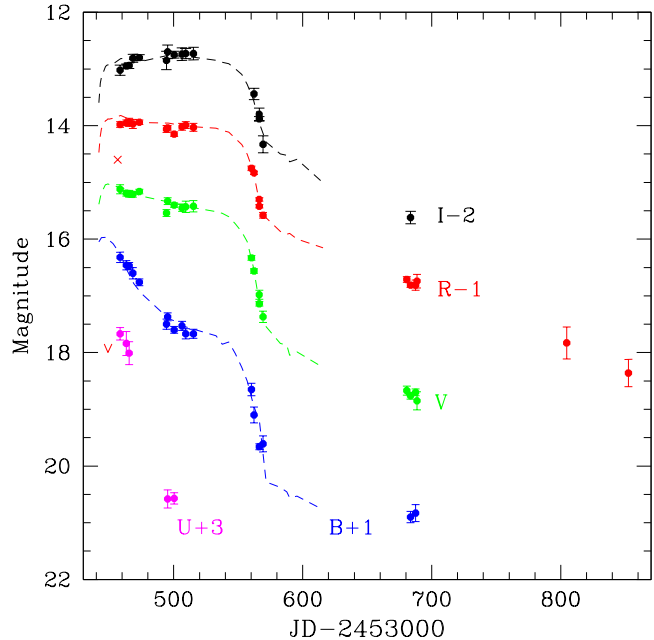}
\caption{$UBVRI$ light curves of SN 2005ay. Dots show
our data, cross is for the discovery magnitude by Rich (2005),
the ``v'' mark is the upper limit from Yamaoka \& Itagaki (2005). 
The dashed lines are the light curves of SN 1999gi.}
\end{figure}
{\bf SN 2005ay}. We began to monitor this SN the day
after discovery, on 2005 March 28, and the last image was obtained
on 2006 April 26, more than a year after explosion.   
The light curves are shown in Fig. 3,
where we also plotted the magnitude estimate by Kloehr on
discovery images and the upper limit
reported by Yamaoka and Itagaki (2005)\nocite{Yam05}.
The light curves appear typical for type IIP SN, this being evident from
comparison with the light curves of SN 1999gi (Leonard et al. 
 2002b)\nocite{Leo02}. 
We shifted the light curves of SN 1999gi along the time axis to
reach the best agreement for the stage of rapid decline after the plateau,
and it is clear that the plateau phase was shorter
for SN 2005ay ($\sim$ 10$\%$ lower plateau duration). 
At this stage the brightness in the $R$ band was constant, but we can
notice slight brightening in $I$ towards the middle of the plateau,
and a decline of luminosity in $V$ at a rate of about 0.006 mag/day.
In the $B$ and $U$ bands the luminosity decreased steadily, and the rate
of decline in $U$ amounted to 0.08 mag/day. After the plateau phase,
the light curves show a rapid drop of about 2 mag, followed
by a linear decline in magnitude. We can estimate the rate of decline
only in the $R$ band, and it is very close to the standard value
of 0.01 mag/day, similar to the radioactive decay of $^{56}$Co.     

We estimate the following magnitudes at maximum light, which was 
reached at the onset of the plateau in all bands except $I$:
$U_{max}=14.7, B_{max}=15.3, V_{max}=15.2, R_{max}=15.0, I_{max}=14.7$. 
The light of SN 2005ay suffered minor extinction in the Galaxy: 
$A_B=0.09$ according to Schlegel et al. (1998)\nocite{Sch98}. We estimate the  
total extinction by comparing the colour curves of SN 2005ay with 
those for SN 1999gi. The comparison shows that in the first 20 days 
after discovery the $(B-V)$ colour of SN 2005ay is very close to 
SN 1999gi, being only 0.04 mag bluer. But at the phase 35-60 days 
SN 2005ay is significantly bluer than SN 1999gi, by about 0.13 mag.
 Thus there is an intrinsic difference in the observed
 colour curves of
these SNe, and we suggest that the colour excess for SN 2005ay was
about 0.1 mag less than for SN 1999gi. As the total colour excess was 
estimated as $E(B-V)=0.21\pm0.09$ mag for SN 1999gi (Leonard et al. 2002), 
 we propose 
$E(B-V)\approx 0.1$ mag for SN 2005ay. Using this estimate
 the corrected-intrinsic $(B-V)$ colour curves are illustrated
 in Fig. 6(bottom panel). Adopting a distance modulus
$\mu =30.82$ from the ``LEDA'' database\footnote{http://leda.univ-lyon1.fr/}, 
we derive absolute magnitudes at a maximum of $M_B \simeq M_V \simeq -15.9$.
This is about 0.8 mag fainter than the mean absolute magnitude for
SNe IIP from Richardson et al. (2002)\nocite{Rich02} recalculated to 
the value of $H_0=70$ km s$^{-1}$ Mpc$^{-1}$.  We note here that 
 NGC 3938 belongs to the $Ursa~ Major$ group. The spiral galaxy NGC 3982, which
 is also part of the group, has a Cepheid calibrated distance modulus
 of 31.71 (Saha et al. 2001)\nocite{Saha01}.If assuming the two 
 galaxies have this same distance, then SN 2005ay would
 be intrinsically more luminous, which would increase the synthesized 
 $^{56}$Ni mass by a factor of $\sim$2.26 (see Sect. 4). The front-to-back 
 effects on the relative distances in the group remain uncertain. 
 The significance of the difference
 in recession velocity of the two galaxies ($\sim$0.0037 and $\sim$0.0027 
 for NGC 3982 and NGC 3938, 
 respectively; LEDA database) also remains uncertain.   
\medskip
\noindent

{\bf SN 2005cs}. We started observations of this object on July 3,
5 days after discovery, and continued monitoring until 2006 April 24.
The light curves are shown in Fig. 4, where we also plotted 
data from amateur astronomers reported on the Astrosurf 
site\footnote{http://www.astrosurf.com/snweb2/2005/05cs/05csMeas.htm},
obtained before our first data point. For comparison we
plotted the light curves for a typical SN IIP 1999em (Leonard et al. 2002a;
 Elmhamdi et al. 2003a; Hamuy et al 2001).
SN 2005cs is certainly of type IIP, but the light curves 
are noticeably different from those of SN 2005ay. 
 The early magnitudes from amateurs, taken by different and independent 
 obsevers, suggest the presence
 of a narrow peak before the onset of the plateau in the $V$ 
and $R$ bands at JD 2453551-552, but we cannot exclude that it is due to the 
errors of their data. Nevertheless, we can conclude that the fast
brightening finished at that epoch, and that maximum light in the $B$
band was reached on JD 2453553 with $B_{max}=14.5$ mag. 
In the $V$ and $R$ bands the luminosity declined slowly until 
JD 2453580, and then a slight brightening is seen during
 the plateau phase. 
At the same epoch the rate of decline in $B$ slowed down.
In the $I$ band, the brightness of
 the SN increased from the onset of the plateau stage and maximum 
light in $I$ and $R$ was reached on JD 2453620, with 
$R_{max}=14.15, I_{max}=13.80$. In the $V$ band, SN 2005cs
was brightest at a very early epoch, with $V_{max}=14.3$.
A similar behavior was observed for SN 1999em at the plateau stage,
but the increase in brightness was more pronounced for SN 2005cs.
The plateau stage lasted until about day JD 2453660,
and then the light curves display a steep decline.
 The luminosity dropped by about 2.8 mag,
the decline was significantly greater than for SN 1999em,
as suggested by Li et al. (2006).
The final exponential tail started at about JD 2453680, and 
the decline rate in $R$ was $\sim$0.007 mag/day. In the other bands
the quality of our data does not allow to estimate the decline rate
 (see Sect.4 for discussions about late V-photometry and $^{56}$Ni 
 estimates).
\begin{figure}
\includegraphics[height=9cm,width=9cm]{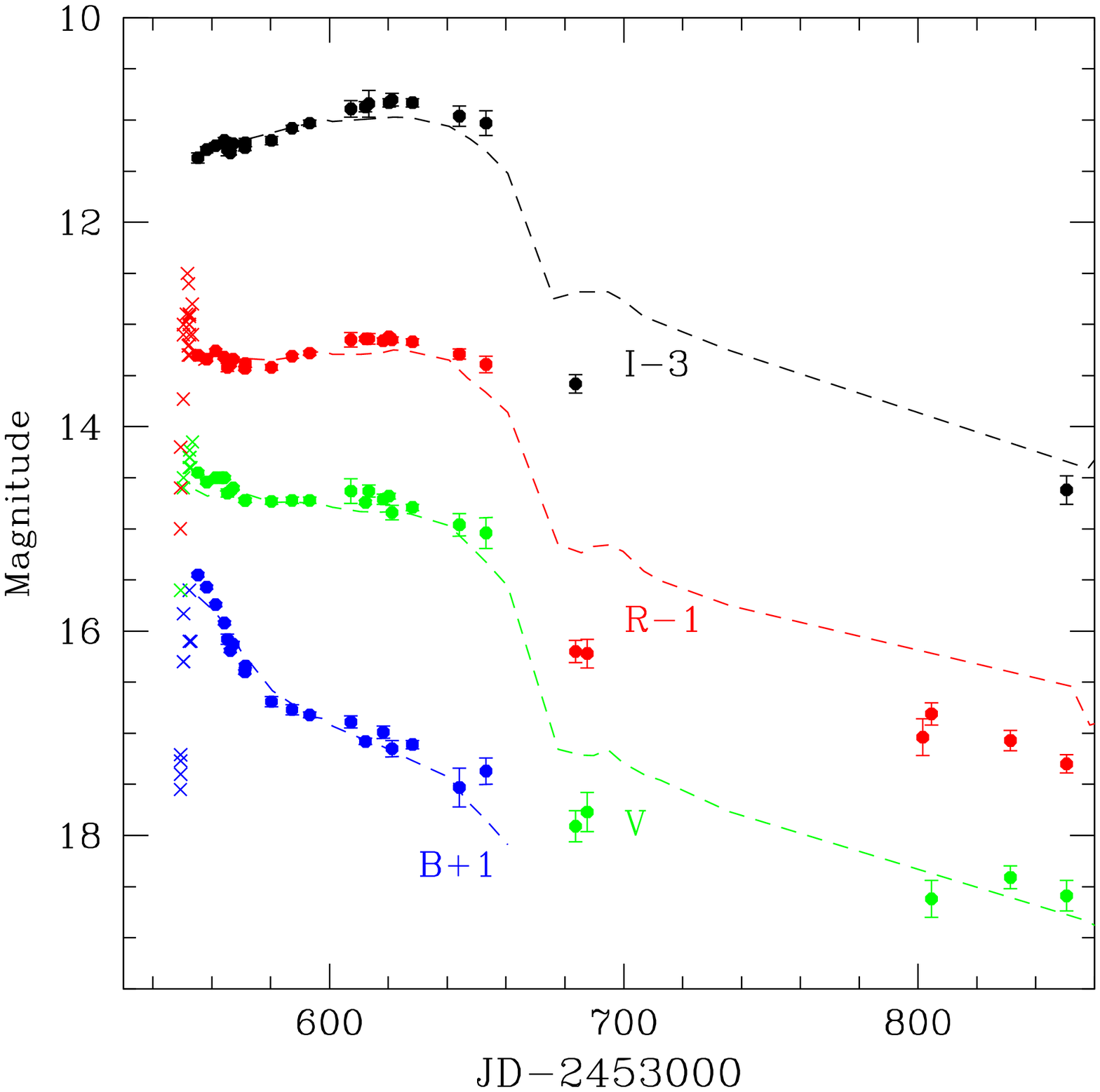}
\caption{$BVRI$ light curves of SN 2005cs. Dots show
our data, crosses are for the observations of amateur astronomers.
The dashed lines are the light curves of SN 1999em.}
\end{figure}

According to Modjaz et al. (2005) and Li et al. (2006), the most
likely value of total colour excess for SN 2005cs is 
$E(B-V)\approx0.1$. A more detailed view regarding the colour
 excess for SN 2005cs has been presented by Maund et al. (2005)
 The analysis suggests an amount of total reddening of
 $E(B-V)=0.14 \pm 0.2$, which we adopt in our present work.
 The ($B-V$) colour evolution is shown in Fig. 5 (lower panel),
 and compared to SNe 1987A, 1999em, 1999gi and 2005ay.
Taking this value for 
extinction and a distance modulus of $\mu=29.6$ (Feldmeier et al. 
 1997\nocite{Feld97}; Maund et al. 2005),
 SN 2005cs is
 an underluminous type IIP SN, more than a magnitude fainter
than the mean IIP luminosity (Richardson et al. 2002) (see details in Sect.4). 

The spectrum of SN 2005cs obtained on July 6 (JD 2453558.37)
is presented in Fig. 6. The H$\beta$ and H$\gamma$ features are
 prominent with typical P-Cygni profiles, and 
Fe II absorption lines are also evident, with observed absorption 
minima at 4955\AA\ and 5094\AA. From the wavelengths of 
H$\beta$ and Fe II $\lambda$5169 absorption minima, we can derive
expansion velocities of 5575 and 4984 km s$^{-1}$, taking 
 600 km s$^{-1}$ as the 
recession velocity of M~51. A spectrum analysis was made by means
 of the synthetic code SYNOW in section 4.2. We see that 
the expansion has slowed down by
about 2000 km s$^{-1}$ since the observation by Modjaz et al. (2005).
These velocities are
significantly lower than the values for SN 1999em (Leonard et al. 2002a)
at similar epochs. 

\section{Main physical properties}

\subsection{Estimate of ejected $^{56}$Ni mass}
 \begin{figure}
\includegraphics[height=10cm,width=9cm]{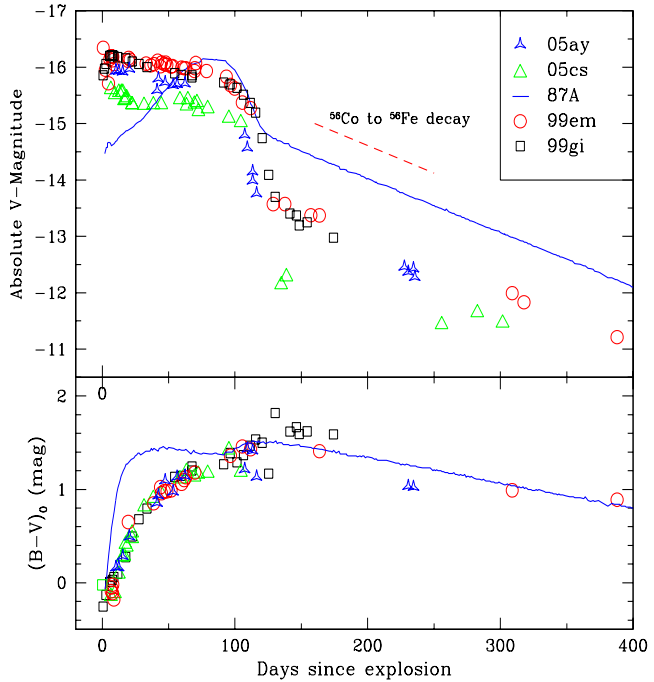}
\caption{ Upper panel: the absolute V-light curves of SNe 2005ay (shorter 
 distance case) and 
 2005cs, compared to SNe 1987A ($D=50~ kpc, A_V^{tot}=0.6$ mag), 
 1999em ($D=8.8 ~Mpc; A_V^{tot}=0.31$ mag) 
 and 1999gi ($D=10.91 ~Mpc; A_V^{tot}=0.65$ mag). For SN 2005ay we plot the
 shorter distance case. Lower
 panel: the ``$(B-V)$'' intrinsic colour evolution of the SNe sample.}
\end{figure}

 The absolute V-band light curves of SNe 2005ay and 2005cs are illustrated
 in Figure 5(upper panel). For SN 2005ay, the shorter distance is used 
 in the plot.
 The light curves of the peculiar SN 1987A
 and SNe IIP 1999em and 1999gi are displayed for comparison.
 For the $^{56}$Ni mass determination the total extinction and distance 
 estimates are crucial. The adopted parameters for the well studied 
 SNe 1987A, 1999em and 1999gi are listed on the caption of Figure 5.
 The assumed explosion time also affects the  $^{56}$Ni mass estimates.
 For SN 2005cs the explosion time is estimated to be June 27.3 
($JD 2453548.8$), or earlier by as much as 1.5 days. This is based on the 
 early rise in the light curves, especially in the B-band, from our
 data combined with the online Astronomers 
 data\footnote{http://www.astrosurf.com/snweb2/2005/05cs/05csMeas.htm}. 
 Following 
 the discussion by Maund et al. (2005), we adopt a total colour excess and 
 distance to the host galaxy of $E(B-V)=0.14 \pm0.2$ and 8.4 Mpc, 
 respectively. 

 The situation of SN 2005ay is less clear. On the one hand, 
 the very early photometry of the rise is lacking, 
 although our earliest data indicate a slight trend of increase 
 in the apparent luminosity
 in  the $R$ and $I$ bands. On the other hand, the similarity to the
 SN 1999gi light curves
 is evident, both on the plateau and on the transition phase to reach
 the radioactive tail (Figure 3).
  The SN 1999gi light curve shape could be used to fit SN 2005ay data and 
 hence obtain an estimate of the explosion epoch. This leads to
  $t_{0} \simeq$ March $ 13~ (JD=2453443)$. However a CCD image taken
 on March 19.621 ($JD=2453449.62$) showed nothing at the location of the 
 supernova (Yamaoka \& Itagaki 2005). This analysis therefore demonstrates 
 the $incorrectness$ 
 of estimating explosion times based on light curve fits, even for seemingly
 similar events within homogeneous class of supernovae such as type 
 IIP. A direct consequence of this reasoning is that SN 2005ay has a short
 plateau compared to SN 1999gi, implying possible differences in the 
 physical parameters, especially the ejecta mass, athough the velocity
 behaviour and the bolometric plateau luminosity also must be
  considered.

 The early evolution of the intrinsic $(B-V)$ colour can provide
 further constraints on the explosion time. Assuming a total colour excess
 of $E(B-V)=0.1$ towards SN 2005ay, we estimate the explosion time to be
 $t_{0} \simeq$ March $ 23~ (JD=2453453)$, with an uncertainty of 2 days.
 This is based on the assumption of a similar photospheric-phase 
 temperature evolution
 behaviour in type IIP objects, which is related to the adiabatic cooling
 of the radiation trapped in the expanding envelope (Schmidt et al. 
 1992\nocite{Schm92}; Elmhamdi et al. 2003b\nocite{Elmha03b}). 
 The difference to SN 1987A is indeed due
 to a faster cooling due to the smaller presupernova radius (i.e. the blue
 supergiant nature of the progenitor star).  
 The standard reddening laws of Cardelli et al. (1989)\nocite{Card89} 
 are adopted. Figure 5
 (bottom panel) displayed the intrinsic colour curves of the five SNe.
 During the first 100 days the similarity between the SNe, except SN 1987A,
 gives us confidence in the assumed parameters and hence the reliability
 of the absolute V-light curves that will be used later for the estimation
 of $^{56}$Ni mass. The main parameters of the two events are summarized
 in Table 6. 
  
 The absolute V-light curves of SNe 2005ay and 2005cs, Figure 5 (top panel), 
 highlight the different phases of evolution compared to SNe 1987A, 
 1999em and 1999gi, during the first 300 days. At early times SN 2005cs 
  shows a brief peak followed
 by a settling onto the plateau phase, around day 18 after explosion, in which
 the magnitude remains almost constant, accounting for the cooling 
 and the expansion balance. An absolute magnitude of 
 $M_v \simeq -15.33$ is measured from a least
 squares fit of the plateau data (from +18d to +105d). Later the light
 curve exhibits a steep decline, $\sim$ 2.8 mag in $\sim$ 30 days, signaling 
 the onset of the nebular phase and the start of the exponential tail. The 
 radioactive decay phase is clearly not well sampled, nevertheless one may
 remark a slower decay compared to the $^{56}$Co slope (dashed line in Fig. 5).
 Similar flattening has been reported in intrinsically fainter objects 
 (Elmhamdi et al. 2003a). In the cited work, it has also been noted 
 that the duration of this 
 flattening period, seen also in other bands, might correlate with the
 synthesized $^{56}$Ni mass, with lower $^{56}$Ni mass objects 
 having a longer duration
 flattening. The last two photometric points show a decline rate close 
 to the  $^{56}$Co decay. Therefore we rely on these late 
 data to recover
 the amount of the ejected $^{56}$Ni. Adopting a $^{56}$Ni mass of
 0.075 $M_{\odot}$ for SN 1987A (Catchpole et al. 1988\nocite{Catch88}; 
 Bouchet \& Danziger 1993\nocite{Bouch93})
 , a direct shift of the light curve tail of SN 1987A to fit late
 data of SN 2005cs indicates a synthesized amount of $^{56}$Ni of 
 $\sim$0.017 $M_{\odot}$. 

 The $M_V$ photometry observations of SN 2005ay indicate
 an early decline from a short maximum (lasting about 17 days 
  with $M_V^{max}$ $\simeq$ -15.94 for the shorter distance and -16.83 
 for the larger one), reaching the plateau phase with 
 a measured mean absolute magnitude of about 
 $M_V$ $\simeq$ -15.72 and -16.61 for the shorter and longer distances,
 respectively. Our photometry is poorly sampled at the end of 
 the plateau phase, however the data at the transition phase 
 plateau-tail seem to indicate a shorter plateau phase compared to 
 SNe 2005cs and 1999em. The observed data, around day 230 after 
 explosion, span about 8 days of observations. These late
 data are used as above to recover the 
 $^{56}$Ni mass (i.e. best fit with SN 1987A radioactive tail). This 
 method leads to $\sim$0.023 $M_{\odot}$ for the shorter distance
 and $\sim$0.051 $M_{\odot}$ for the larger distance.  
    
 As a further check of the derived $^{56}$Ni masses of the two events, we make
 use of the well established correlation between the $^{56}$Ni mass 
 and the absolute $M_V$ at the plateau phase (Elmhamdi et al. 2003b; Hamuy
 2003\nocite{Ham03}).
 Adopting the previous estimates of $M_V$ during the plateau phase
 , i.e. $M_V$ $\simeq$ -15.33 for SN 2005cs and $M_V$ $\simeq$ -15.72; -16.61 
 for SN 2005ay , the method gives $^{56}$Ni masses of $\sim$0.018 
 $M_{\odot}$ for SN 2005cs and $\sim$0.026 $M_{\odot}$ for the shorter distance
  and $\sim$0.065 $M_{\odot}$ for the larger distance for SN 2005ay.
 
 It is reassuring that the two methods give consistent results 
 and indicate rather low
 ejected $^{56}$Ni masses for SN 2005cs and for SN 2005ay if the shorter 
 distance obtains. A higher amount for SN 2005ay obtains  
 for the larger distance. This indicates that at least SN 2005cs
 belongs to the faint tail of the luminosity function of 
 type II SNe while the situation for SN 2005ay depends on the adopted
 distance.
\subsection{The progenitor characteristics}
 Making use of the type IIP-analytical model  of Popov (1993), and 
 the hydrodynamic models of Litvinova \& Nadyozhin (1985), it is possible 
 to derive physical parameters, namely energy $(E)$, ejecta mass $(Mej)$ 
 and presupernova radius $(R)$. The duration of the plateau $(tp)$, 
 the absolute 
 V-magnitude at the plateau phase $(Mv)$ and the photospheric velocity $(Vp)$ 
 are the input parameters, usually estimated from the observations. 
 
 As discussed previously, from observations of SNe 2005ay and 2005cs we
 adopt $M_V (plateau)$ $\simeq$ -15.72;-16.61 and $M_V (plateau)$ 
 $\simeq$ -15.33, 
 respectively. The early rise seen in both light curves of the two objects
 give constraints on the duration of the plateau (starting later than
 2 weeks after explosion). We adopt a range of $tp=80-90$ days
 for the two events, although SN 2005ay clearly has a shorter plateau
 duration than SN 2005cs. This range represents our estimate
 of the uncertainty associated with the determination of this parameter.
 We note that more precise constraints 
 on the plateau duration
 should come from an integrated bolometric light curve shape. The obvious
 observed initial declines in $U$ and $B$ bands would affect the
 resulting constructed bolometric light curves, indicating 
 a shorter plateau phase than is seen in the V-band light curve
 (for expample SN 1999em; Elmhamdi et al. 2003a)
\begin{table*}
\caption{The SNe progenitor properties}\vskip2mm
\begin{tabular}{ccclcccl}
\hline
\multicolumn{4}{c}{\hskip4cm SN 2005ay} & \multicolumn{4}{c}{SN 2005cs}\\
\hline
\hskip1cm $t_{plateau}$ & $Mej$ & $E$ & $R$ &  &$Mej$&$E$ & $R$\\
\hskip1cm ($days$) & ($M_\odot$) & (10$^{51}$ erg/s) & ($R_\odot$)&&$(M_\odot$) & (10$^{51}$ erg/s) &($R_\odot$) \\
\hline
\hskip-2cm Popov :\\   
\hskip1cm 80 &4.55$-$5.1& 0.24$-$0.5 & 393.5$-$582 && 4.33 & 0.17 & 331  \\
\hskip1cm 90 &7.28$-$8.18& 0.38$-$0.8 & 310$-$460   && 6.93 & 0.28 & 261   \\
\hline
Litvinova \& Nadyozhin :\\ 
\hskip1cm 80 &9.12$-$10.4& 0.41$-$0.8 & 202.4$-$278 &&8.61  &0.3  &176    \\
\hskip1cm 90 &12.8$-$14.6& 0.53$-$1.08 & 178.4$-$245 &&12.13 &0.39  & 155   \\
\hline
\hskip-0.5cmEjected $^{56}$Ni mass:\\
&&$0.023-0.026$$~M_\odot$ (Shorter distance)&&&&$0.017-0.018$$~M_\odot$&\\
&&$0.051-0.065$$~M_\odot$ (Larger distance)&&&&&\\
\hline
\end{tabular} \\ 
\footnotesize
\hskip-5.5cm\emph{\rm SN 2005ay: M$_V(Plateau)=-15.72$ $\&$ V$_{phot}=2980~ km~ s^{-1}$ (Shorter distance);\\$~~~~~~~~~~~~~~~~~~~~~$M$_V(Plateau)=-16.61$ $\&$ V$_{phot}=4070~ km~ s^{-1}$ (Larger distance) }\\ 
\hskip-5.5cm\emph{\rm SN 2005cs: M$_V(Plateau)=-15.33$ $\&$ V$_{phot}=2600~ km~ s^{-1}$}\\ 
\normalsize
\end{table*}
   
 The third input parameter is the representative photospheric velocity
 during the plateau phase. In principal this should be estimated from
 weak lines in the observed spectra during the plateau epoch.
 Unfortunately we lack such observations for the two SNe. However,
 an available spectrum of SN 2005cs is observed around 9.5 days 
 since explosion.
 In order to analyse this spectrum we use the parametrized supernova synthetic 
 spectra code SYNOW. Details of the fitting procedure and parameters 
 can be found elsewhere (see for example Elmhamdi et al. 2006\nocite{Elmha06}).
 Figure 6 showed the observed spectrum, after corrections
 for reddening and host galaxy recession velocity, compared to 
 the best fit synthetic spectrum that has a blackbody temperature
 of T$_{bb}=$ 14500 K, a photospheric velocity of 
 V$_{phot}=$ 5000 km s$^{-1}$,
 and including 7 ions, namely: H I, Fe II, Ca II, Mg II, Mg I, Sc II and Ni II.
 Synthetic line profile features are labeled by the designation of the ion. 
 The observed spectrum has been smoothed with a box size of 
 6(in pixels). 
 The modeling has been undertaken adopting an exponential radial dependence
 of the optical depth (and density), $\propto exp(-v/v_e)$, with 
 an $e-$folding velocity $v_e = 2000$ km s$^{-1}$.
 The match is satisfactory as the main features are accounted for by 
 the synthetic spectrum. On one hand, the good fit with Fe II, 
 especially Fe II 5169 \AA,
 gives us confidence in the deduced photospheric velocity. On the other 
 hand, the Balmer H I features, especially H$\beta$, are clearly 
 identified, and found to have a velocity V$_{min}(HI)=$ 5500 
 km s$^{-1}$.  
 The supernova has a low velocity behaviour with clearly narrow features. 
 Indeed, at similar phases the well observed SNe II 1987A and 1999em 
 had velocities about 4000 km s$^{-1}$ higher.
 We note here that an expansion velocity of about 7500 km s$^{-1}$ has been 
 derived from the minimum of the H$\beta$ from the published 
 SN 2005cs CFA-spectrum 3 days after explosion 
 (Modjaz et al. 2005). Again, at a similar age, H I Balmer 
 lines in SN 1987A had an expansion velocity of the order of 16000 km s$^{-1}$.
 We note also that in their spectrum there occur features reminiscent
 of what was ascribed to N II lines in SN 1999em (Baron et al. 2000).
 Our SYNOW test modeling of the CFA-spectrum suggests that N II lines
 could be the correct interpretation also for SN 2005cs. 
    
 For our purpose, i.e. analytical models, we need an appropriate 
 estimate of the photospheric velocity during the photospheric phase. 
 We propose two different methods, based on two established results
 in type IIP SNe. First, the correlation $M_V(50d).vs.V_{phot}(50d)$ 
 that relates the photospheric velocity to the V-absolute magnitude, 
 both evaluated at day 50 ($V_{phot}(50d) \propto 
 L_{plateau}^{0.33 (\pm 0.04)}$, Hamuy 2003).
 Second, the tight relation found between the photospheric velocity at 50 days
 determined from the Fe II features and the velocity at a given time $t$.
 The best fit equation, based on a type IIP sample study, is: 
 $V_{phot}(50d) = V(t)*(t/50)^{0.464 \pm 0.017}$ (Nugent et al. 
 2006)\nocite{Nug06}.
\begin{figure}
\includegraphics[height=9cm,width=9cm]{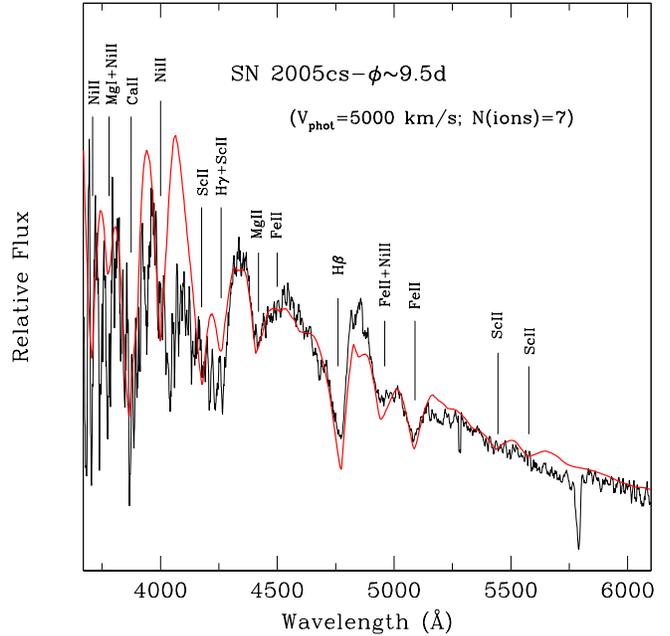}
\caption{Spectrum of SN 2005cs obtained on July 6, 2005 
(JD 2453558.37) at S100, compared with the SYNOW synthetic 
 spectrum (thin line). Lines that are responsible for the most 
 conspicuous features are indicated.}
\end{figure}

 For SN 2005cs, we obtain estimates of $V_{phot}(50d)$ of $\sim$ 2600 and 
 2315  km s$^{-1}$ applying the two described methods.     
 For SN 2005ay, since we do not have an early estimate of the Fe II velocities
 we rely only on the photospheric velocity
 recovered from the first method. We find $V_{phot}(50d) \simeq 2980$ 
 km s$^{-1}$ for the shorter distance and $\simeq 4070$ km s$^{-1}$ for the
 larger distance.
 The resulting parameters derived using both sets of analytical models 
 are summarized in Table 7 for both SNe. These estimates, although inevitably 
 suffering from various uncertainties, indicate that the two SNe, in particular
 SN 2005cs,
 represent examples where all three physical parameters, velocity, energy
 and $^{56}$Ni mass are lower than average, a correlation not always
 observed in SNe IIP.
 Furthermore, recent results using pre-explosion imaging of SN 2005cs
 provide constraints on the progenitor main sequence mass
 (Maund et al. 2005).   
 The authors found that the initial mass of the star is constrained to be
 $M_{ZAMS}=9^{+3}_{-2}~ M_\odot$, which is consistent within the
 uncertainties with the results of the present work.  
 
\medskip
\medskip
\medskip
\medskip
We conclude that both  type IIP SNe 2005ay with the shorter adopted 
 distance, and 2005cs
 have lower than average luminosities, with SN 2005ay being slightly brighter 
and the shape of its light curves being more typical of SNe IIP. Adopting
 the larger distance possibility increases the luminosities and consequently
 the amount of the ejected $^{56}$Ni.
SN 2005cs is distinguished by a slight brightening in 
the $V,R,I$ bands during the second half of the plateau stage,
 and by a low expansion
velocity. Combining these results with the derived data on the 
 progenitor stars, we suggest that these events resulted from the outburst of
 supergiants towards the lower end of the range of accepted progenitor masses
  with low explosion energies and low $^{56}$Ni yields.

\medskip

{\bf Acknowledgments:}
This research has made use of the Lyon-Meudon Extragalactic Database
(LEDA) and of the Canadian Astronomy Data Centre. The authors are
grateful to N.N.Pavlyuk and I.M.Volkov, who carried out some
observations, to S.V.Antipin and K.Sokolovskij for help in the
observations and to O.S.Bartunov for help with computing facilities. 
The work of D.Yu.Tsvetkov was partly supported by RFBR grant 05-02-17480.
 A. Elmhamdi
 is grateful to David Branch for the use of the SYNOW code. I.J. 
 Danziger was supported in this work by a MIUR-Cofin grant from the Italian
 Ministry of Education. We thank the referee, M. Della Valle, for the useful
 comments.

\end{document}